\documentclass[aps,prl,twocolumn,superscriptaddress,showpacs,showkeys]{revtex4}
\usepackage[]{graphicx}
\begin{document}

\title{Singular point characterization in microscopic flows}
\author{Giorgio Volpe}
\affiliation{ICFO - Institut de Ciencies Fotoniques, Mediterranean Technology Park, 08860, Castelldefels (Barcelona), Spain}
\author{Giovanni Volpe}
\affiliation{ICFO - Institut de Ciencies Fotoniques, Mediterranean Technology Park, 08860, Castelldefels (Barcelona), Spain}
\author{Dmitri Petrov}
\affiliation{ICFO - Institut de
Ciencies Fotoniques, Mediterranean Technology Park, 08860, Castelldefels (Barcelona), Spain}
\affiliation{ICREA - Instituci\'{o} Catalana de Recerca i Estudis Avan\c{c}ats, 08010, Barcelona, Spain}
\date{\today}

\begin{abstract}
We suggest an approach to microrheology based on optical traps
in order to measure fluid fluxes around singular points of fluid flows.
We experimentally demonstrate this technique,
applying it to the characterization of controlled flows produced by
a set of birefringent spheres spinning due to the transfer of light angular momentum.
Unlike the previous techniques, this method is
able to distinguish between a singular point in a
complex flow and the absence of flow at all; furthermore it
permits us to characterize the stability of the singular point.
\end{abstract}

\maketitle

The experimental characterization of fluid flows in
micro-environments is important both from a fundamental point of
view and from an applied one, since for many applications it is
required to assess the performance of microfluidic structures,
such as lab-on-a-chip devices \cite{Knight2002}. Carrying out this
kind of measurements can be extremely challenging. In particular,
due to the small size of these environments, wall effects can not
be neglected \cite{Zhu2001}. Additional difficulties arise
studying biological fluids because of their complex rheological
properties.
\newline
In the cases of practical interest the flow is strongly viscous
(creeping motions or Stokes flows) and a low Reynolds number
regime can be assumed \cite{HappelBrenner}. Since the creeping motion regime is a particular
case of a laminar regime, it is possible to univocally define a
time-independent pressure and velocity field. Therefore,
for each position a well-defined drag force acts on a particle immersed in the fluid flow.
Ideally micro-flow sensors should be able to monitor the
streamlines in real time and in the least invasive way. One common
method to achieve this goal is to measure the drag force field
acting on a probe particle, resorting to statistical criteria of
analysis because of the intrinsic presence of Brownian motion.
\newline
Recently optically trapped microscopic particles have been
proposed as flow sensors
\cite{Nemet2002,Bishop2004,Knoner2005,Pesce2006,DiLeonardo2006}.
An optical trap enables the confinement of micron sized objects: a
high numerical aperture objective lens is used to tightly focus an
optical beam and to produce a force sufficient to hold dielectric
objects ranging in size from 100Õs of nanometers to 10Õs of
microns \cite{Neuman2004}. In \cite{Nemet2002} an oscillating
optically trapped probe is used to map the two-dimensional flow
past a microscopic wedge. In \cite{Bishop2004} and
\cite{Knoner2005} a stress microviscometer is presented: it
generates and measures microscopic fluid velocity fields,
monitoring the probe particle displacement, which is directly
converted into velocity field values, through digital video
microscopy. A further improvement was achieved in
\cite{DiLeonardo2006} by using multiple holographic optical traps
in order to parallelize the technique: an array of micro-probes
can be simultaneously trapped and used to map out the streamlines
in a microfluidic device. All these techniques apply a Photonic
Force Microscope (PFM) approach
\cite{Ghislain1993,Florin1997,Rohrbach2002,BergSorensen2004} to
the flow measurement: the fluid velocity at the trap location is
obtained by monitoring the probe displacements resulting from the
balance between the trapping and drag forces.
\newline
All the techniques described above present a major drawback: they
interpret the experimental results assuming that the flow can be
described by a set of parallel streamlines. When this is the case,
the drag force acting on a probe around a specific point in the
flow is well described by a constant value, and this is the case
considered by the current optical trap methods. However, in the
drag force-field there may be singular points as well. In these
points the flow and, hence, the drag force are null, but not in
its surrounding. The question, which arises naturally, is how to
characterize this case. Under the assumption of a steady
incompressible flow, a zero body force and low Reynolds numbers,
the local fluid motion satisfies the quasi-static Stokes equations
\cite{HappelBrenner}: $\eta\nabla^2 \mathbf{v} = \nabla p$,
$\nabla \cdot \mathbf{v} = 0$, where $\mathbf{v}$ is the velocity,
$p$ is the pressure and $\eta$ the dynamic viscosity. Since the
fluid is incompressible and there are no sources or sinks, there
can be only two kinds of singular points \cite{Hirsch}: (1)
saddles (unstable) at the meeting point between two opposite
flowing streamlines or (2) centra (stable) in presence of a
non-zero curl. As we show below, the characterization of the flow
near these points can be achieved by studying the statistical
properties of the Brownian motion of the probe.
\newline
The knowledge of the flow-field near a singular point is of
relevance for fundamental physical studies as well as for
engineering applications. The mixing of fluids flowing through
microchannels is important for many chemical application; a
reduction of the mixing length can be achieved by the generation
of a transverse flow \cite{Stroock2002}, case in which the
formation of singular points is inevitable.  In biological systems
creeping motions take place in small blood or linfatic vessel or
at the interface between tissues and prothesis or artificial
organs \cite{Belanger}. In the presence of slow flows, macrophage
adhesion becomes more probable, increasing the possibilities of
inflammation. Furthermore, the growth rate of thrombi, due to
platelet aggregation, is also determined by the characteristics of
the flow around it: for example, the presence of stable
equilibrium points, such as centra, helps their formation and
growth \cite{Liu2002}. Flow measurements at the microscale can
help to diagnose pathologies and to guide the design of
bio-materials and nano-devices for diagnostic or therapeutic
goals.

In this letter we extend the PFM-approach to microrheology in
order to measure and characterize fluid fluxes in the proximity of
singular points. The concept is to monitor the position of an
optically trapped probe in order to locally characterize the drag
force field as a generic function of the space coordinates up to
the first order in its Taylor expansion around the probe position
\cite{Giorgio2007b}. This technique permits us to distinguish
between a singular point in a complex flow and the absence of flow
at all. Furthermore our approach allows one to determine the
stability of these singular points, which can be relevant for
applications.

In the following we will consider the drag force field produced by
a bidimensional laminar regime. Since in microfluidic and
lab-on-a-chip devices a planar geometry is generally assumed, this
is the most useful case for applications. Nevertheless, if needed,
our approach can be generalized to the three-dimensional case.

Experimentally, we generated these three kinds of fluid flow -
namely a set of parallel streamlines, a saddle or a centrum -
using solid spheres made of a birefringent material (Calcium
Vaterite Crystals (CVC) spheres, radius $R = 1.5 \pm 0.2 \mu m$
\cite{Bishop2004}), which can be made spin due to the transfer of
orbital angular momentum of light. They are all-optically
controlled, i.e. their position can be controlled by an optical
trap and their spinning state can be controlled through the
polarization state of the light. The angular velocity of the CVC
sphere is given by $\omega = \Delta\sigma \lambda P / 16 \pi^2
\eta R^3 c$, where $\Delta\sigma$ is the change of the light spin
momentum due to the scattering on the probe, $P$ is the beam
power, $\eta$ is the medium viscosity, $c$ is the speed of light,
and $\lambda$ is the wavelength of the optical field. In the
experimental realization up to four CVC spheres were optically
trapped in water and put into rotation using four steerable
$1064\,nm$ beams from a Nd:YAG laser with controllable
polarization - to control the direction of the rotation - and
power - to control the rotation rate. A probe polystyrene sphere
(radius $r = 500 \, nm$) was held by an optical trap produced by a
$632\,nm$ linearly polarized beam. The stiffness of this trap was
adjustable through the power. The forward scattered light of this
latter beam served to track the probe position using a position
sensor based on a quadrant-photodetector (QPD).

Assuming no slip at the particle surface, the quasi-Stokes
equation leads to the following solution for the flow velocity
near a single spinning sphere \cite{HappelBrenner,Knoner2005}
\begin{equation}\label{Spinning flow}
    \mathbf{v} = \omega \mathbf{e}_{z} \times \mathbf{x}\frac{R^3}{\left\|\mathbf{x}\right\|^3}.
\end{equation}
In each point near the rotating sphere
the streamlines are perpendicular to the plane containing the $z$-axis
(unit vector $\mathbf{e}_z$) and the coordinate vector
$\mathbf{x}$.
\newline
\begin{figure}[tbp]
\includegraphics[width = 8.5cm]{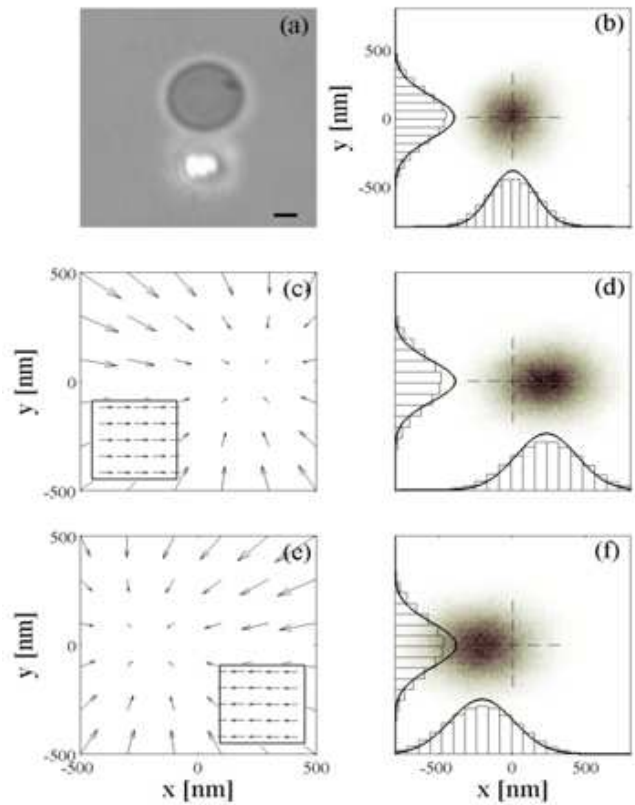}
\caption{(Color online) (a) Image of a trapped probe near a single
spinning sphere. The bar is $1\,\mu m$. (b) 2D Probability Density
Function (PDF) when the spinning sphere is at rest. (c-f) Total
force fields - insets: drag force field contribution -  and 2D PDF
when the sphere is spinning (c-d) counterclockwise and (e-f)
clockwise. For each of the plots (b),(d), and (e) data acquired
during $150 \,s$ with sampling rate $2000\,Hz$ were analyzed.}
\end{figure}
In Fig. 1 the results for probing such a drag force field near a
single spinning sphere are presented. The CVC sphere is put into
rotation at $\omega \approx 38 \,rad/s$. The induced drag force
can be measured through the shift of the equilibrium position of
the probe (Fig. 1(b), 1(d), and 1(f)) and after calibrating the
stiffness of the probe trap ($k = 175 \pm 15 \,fN /\mu m$). The
corresponding force field can be reconstructed (Fig. 1(c) and
1(e)): the forces acting on the probe particle result $40\pm5
\,fN$ (counterclockwise rotation) and $37\pm5\,fN$ (clockwise
rotation). These results are in agreement with
\cite{Bishop2004,Knoner2005}.

In a system with $n$ spinning spheres, the linearity of the
quasi-Stokes equations \cite{HappelBrenner} allows one to use the superposition of
solutions (1) in order to obtain the total flow. Since the
velocity in the equatorial plane is proportional to the reciprocal
square of the distance from the sphere center, the flow velocity vanishes at a distance of
a few sphere radii. Hence, in
first approximation the hydrodynamic interaction between the spinning spheres located far enough from each other
can be neglected.
For $n = 2$ or $n = 4$ we assume the spheres to be
symmetrically displaced at the same distance $\rho$ with respect to the origin $\mathbf{x}_0$
(Fig. 3(a) and 4(a)), so that the origin is a singular point, i.e.
$\mathbf{v(\mathbf{x}_0)} = 0$. The velocity field can be
locally approximated as
\begin{equation}\label{Linearized flow}
    \mathbf{v} = \omega R^3 \mathbf{J}\mathbf{x} = \omega R^3
    \sum\limits_{i=1}^{n} {\mathbf{J}_i}\mathbf{x},
\end{equation}
where $\mathbf{J}$ is the Jacobian matrix of the total velocity
field evaluated in the singular point and $\mathbf{J}_i$ the
Jacobian matrix of the velocity field generated by particle $i$
evaluated in the same point. As expected, since the fluid is
incompressible, $Tr(\mathbf{J})$ is always null. This means that
only (unstable) saddles - with $n=2$, $\det ({\bf J}) =  - 4/\rho
< 0$ - or (stable) centra - with $n = 4$, $\det ({\bf J}) = 2/\rho
> 0$ - can exist.
\newline
\begin{figure}[tbp]
\includegraphics[width = 8.5cm]{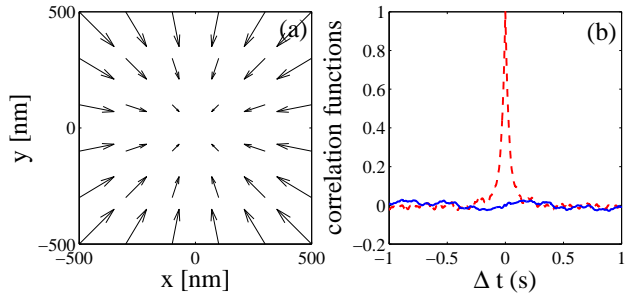}
\caption{(Color online) (a) Force field and (b) auto- (red dashed
line) and cross-correlation (blue solid line) functions when only
the optical force is acting on a $0.5\,\mu m$ radius probe in
absence of drag force field. For plot (b) data  acquired during
$150 \,s$ with sampling rate $2000\,Hz$ were analyzed. }
\end{figure}
In Fig. 2 to 4 the experimental results for such cases are
presented. We followed the data analysis procedure proposed in
\cite{Giorgio2007b}, to which we refer for the full details. The
total force field acting on the probe is given by the sum of the
drag force-field and the restoring force due to the harmonic
trapping potential. The drag force field can be decomposed into a
conservative and a rotational part, while the optical force field
is purely conservative. The cross-correlation function (CCF)
between the $x$ and $y$ position of the probe, calculated as
$CCF(\Delta t) = <x(t) y(t+\Delta t)> - <y(t) x(t+\Delta t)>$,
does not vanish only if the rotational part of the drag force
field is not null. The rotational component of the drag force
field can be obtained from the CCF
\cite{Giovanni2006b,Giorgio2007b}. If the rotational component,
and therefore the CCF, is null, the total force field is purely
conservative, and the conservative component of the drag force
field can be calculated by subtracting the optical potential from
the total potential. We underline that this can only be done if
the total force field is conservative. Some results for an
optically trapped probe in the absence of flow are presented in
Fig. 2: the optical force-field is harmonic (Fig. 2(a)) and it is
purely conservative, as it is shown by the fact that its CCF
function is null (Fig. 2(b)).
\newline
\begin{figure}[tbp]
\includegraphics[width = 8.5cm]{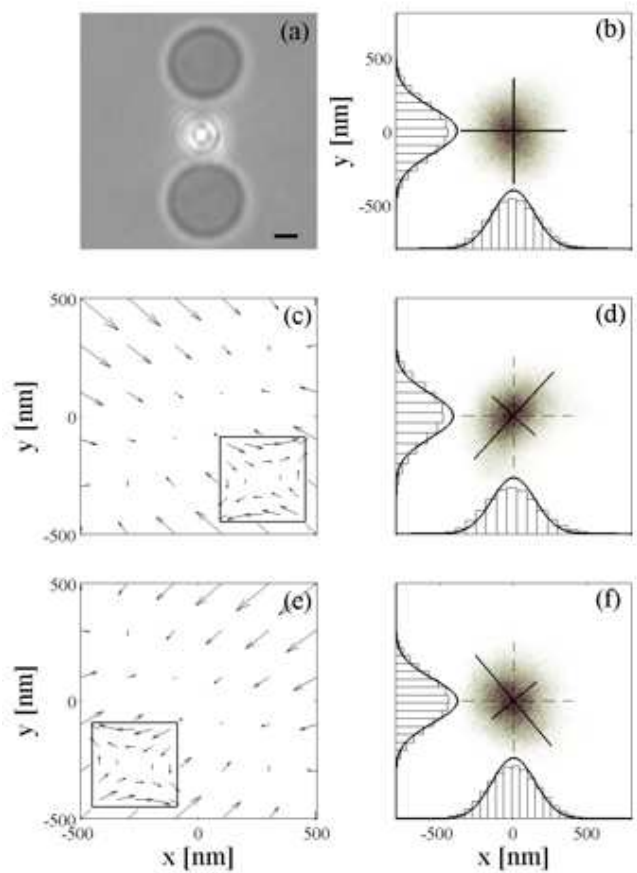}
\caption{ (Color online) (a) Image of a trapped probe between a
couple of spinning spheres. The bar is $1\,\mu m$. (b) 2D
Probability Density Function (PDF) when the spinning sphere is at
rest. (c-f) Total force fields - insets: drag force field
contribution -  and 2D PDF when the sphere is spinning (c-d)
counterclockwise and (e-f) clockwise. For each of the plots
(b),(d), and (e) data acquired during $150 \,s$ with sampling rate
$2000\,Hz$ were analyzed.}
\end{figure}
In Fig. 3 the flow to be probed is generated putting into rotation
($\omega \approx 38 \, rad/s$) two CVCs, symmetrically positioned
with respect to the probe beads (Fig. 3(a)). The position of the
probe is chosen so that there is no shift in the equilibrium
position regardless of the rotation state of the spheres (Fig.
3(b), 3(d), and 3(f)). Since the experimental CCF is found to be
null within the experimental error, we conclude that the
rotational component of the drag force field must be negligible
and we can reconstruct its conservative part by subtracting the
optical restoring force-field from the total force field (Fig.
3(c) and 3(e)). The optical trap produces a symmetric harmonic
potential ($k = 185 \, fN/\mu m$) (Fig. 3(b)). With spinning CVCs
the PDF becomes ellipsoidal (Fig. 3(d) and 3(f)). For a clockwise
rotation (Fig. 3(c-d)), the main axes of the ellipses are oriented
at $40^{\circ}$, the stiffness is $k_{min} = 142 \,fN/\mu m$ along
the main axis and $k_{MAX} = 261\,fN/\mu m$ along the secondary
axis. For a counterclockwise rotation (Fig. 3(e-f)), the main axes
of the ellipses are oriented at $42^{\circ}$, the stiffness is
$k_{min} = 165\pm20\,fN/\mu m$ along the main axis and $k_{MAX} =
211\pm20\,fN/\mu m$ along the secondary axis. The difference are
due to the fact that the rotation rate was observed to be
different in the two directions. The corresponding total force
fields are represented in Fig. 3(c) and 3(e), while the drag force
field contributions are depicted in the insets: they constitute
indeed saddle points.
\newline
\begin{figure}[tbp]
\includegraphics[width = 8.5cm]{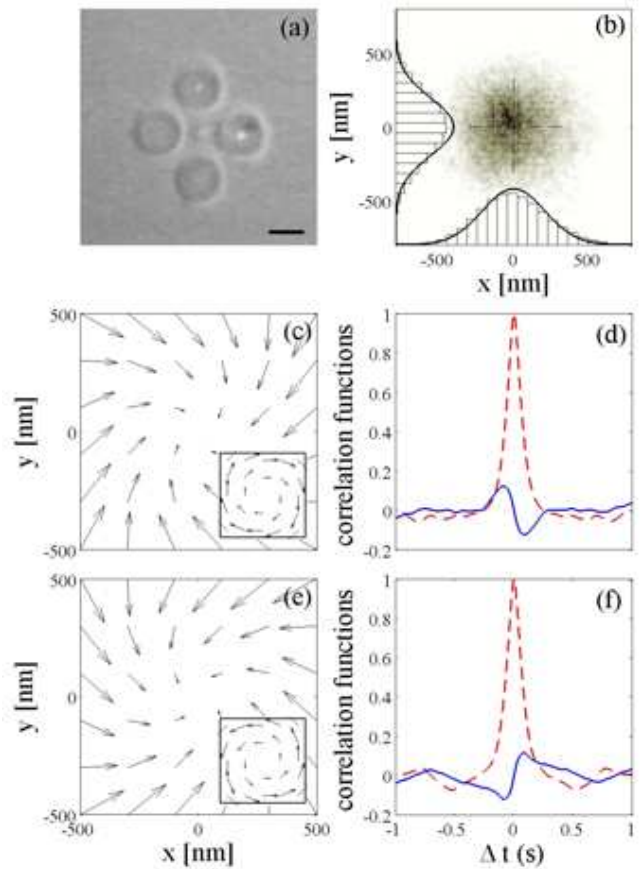}
\caption{(Color online) (a) Image of a trapped probe between four
spinning spheres. The bar is $3\,\mu m$. (b) 2D probability
density function when the spinning sphere is at rest. (c-f) Total
force fields - insets: drag force field contribution - and auto-
(red dashed line) and cross-correlation (blue solid line)
functions when the sphere is spinning (c-d) counterclockwise and
(e-f) clockwise. For each of the plots (b),(d), and (e) ten
datasets acquired during $15 \,s$ with sampling rate $2000\,Hz$
were analyzed and averaged. }
\end{figure}
In Fig. 4 the characterization of a fluid flow near a singular
point of the second kind, a centrum, is presented. This stable
singular point was generated putting into rotation ($\omega
\approx 38\, rad/s$) four birefringent spheres symmetrically
distributed with respect to the probe (Fig. 4(a)) in an optical
trap with stiffness $k = 78 \pm 5\, fN/\mu m$. We can observe that
the PDF of the probe position do not depend on the rotation
direction of the spinning spheres (Fig. 4(b)); however, in this
case the CCF is not null (Fig. 4(d) and 4(f)), showing the
presence of a rotational component of the drag force field. This
component is such that it produces a torque on the probe bead,
whose angular velocity is $\Omega$, as $\gamma \Omega <x^2+y^2> =
5.7 \, fN\mu m$ for the beads rotating clockwise (Fig. 4(c-d)) and
$6.4 \, fN\mu m$ for the case of the beads rotating
counterclockwise (Fig. 4(e-f)). The resulting total and drag force
fields are represented in Fig 4(c) and 4(e), while the drag force
field contributions are depicted in the insets: they constitute,
indeed, a centrum.

In conclusion, we have demonstrated the use of an optically trapped probe for the
characterization of singular points in microscopic flows.
This technique delivers an important contribution
towards a better understanding and optimization
of microfluidic flows in the presence of singular points,
providing both a deeper understanding
of their physics and an experimental approach to their
characterization. Furthermore our
approach allows one to determine the stability of these singular
points, which can be relevant for applications.

\begin{acknowledgments}
The authors acknowledge useful discussions with N. Heckenberg, A.
Bagno, and M. Rub\'{i}. This research was carried out in the
framework of ESF/PESC (Eurocores on Sons), through grant
02-PE-SONS-063-NOMSAN, and with the financial support of the
Spanish Ministry of Education and Science. It was also partially
supported by the Departament d'Universitats, Recerca i Societat de
la Informaci\'{o} and the European Social Fund.
\end{acknowledgments}

\bibliography{biblio}

\end{document}